# Multi-Aperture Fusion of Transformer-Convolutional Network (MFTC-Net) for 3D Medical Image Segmentation and Visualization

Siyavash Shabani[1,*]    Muhammad Sohaib[1]    Sahar A. Mohammed[1]    Bahram Parvin[1,2,3,*]

[1]Department of Electrical and Biomedical Engineering
[2]Department of Immunology and Microbiology, School of Medicine
[3]Pennington Cancer Institute, Reno Nevada
University of Nevada, Reno, NV, USA
*{sshabani, bparvin}@unr.edu

**Abstract**

*Vision Transformers have shown superior performance to the traditional convolutional-based frameworks in many vision applications, including but not limited to the segmentation of 3D medical images. To further advance this area, this study introduces the Multi-Aperture Fusion of Transformer-Convolutional Network (MFTC-Net), which integrates the output of Swin Transformers and their corresponding convolutional blocks using 3D fusion blocks. The Multi-Aperture incorporates each image patch at its original resolutions with its pyramid representation to better preserve minute details. The proposed architecture has demonstrated a score of 89.73±0.04 and 7.31±0.02 for Dice and HD95, respectively, on the Synapse multi-organs dataset—an improvement over the published results. The improved performance also comes with the added benefits of the reduced complexity of approximately 40 million parameters. Our code is available at https://github.com/Siyavashshabani/MFTC-Net*

## 1. Introduction

Vision Transformers (ViT) [3] have demonstrated state-of-the-art performance on major image classification benchmarks [5]. One of the applications of Transformers has been in the area of medical image segmentation [8, 9]. H. Cao and colleagues [8] introduced a novel Transformer-based architecture that replaces the traditional UNet framework for medical image segmentation. Because of the added value of the convolutional blocks, there has been a growing interest in developing methods to integrate representations from the Transformers and convolutional modules in preserving local information. Among these, Swin UNETR [2, 12] employs the Swin technique to construct a UNet-shaped architecture featuring a Transformer-based encoder and a convolutional-based decoder.

In support of the proposed research, it is noteworthy to indicate that the UNet model and its extensions have been examined extensively for biological applications [13, 14]. In basic UNet-CNN models [15-17], the encoder block maps the image to a lower-dimensional (latent) space, which is then reconstructed by the decoder block. Additionally, the skip connections, at each stage of the dimensionality reduction, map the output of the encoder to the decoder blocks to improve medical image segmentation [18]. These blocks excel in extracting local image features, but they are challenged to extract global features and their associations. Some studies have incorporated the attention mechanism [19-21] to enhance performance by improving the extraction of global features.

Although recent advances can grasp local and global contexts, we hypothesized that improved 3D segmentation can benefit from multi-aperture representation for a more accurate representation of surfaces. In this study, we investigated the multi-aperture representation and coupling Transformers and convolutional modules with 3D fusion blocks. Here, multi-aperture refers to overlapping multiscale representations of a region, where each aperture maintains its original resolution. Hence, the surfaces between adjacent objects are not diffused as is the case for multiscale representation. The main contributions of our study are (i) four parallel Swin Transformers modules that couple the pyramid representation with the original resolution; (ii) a 3D fusion block with a squeeze and excitation block and a convolutional block attention module (CBAM) for combining the outputs of each Transformer and its convolutional block; (iii) a loss function that integrates the Dice cross-entropy with a custom distance transform for morphometric consistency; and a reduce model complexity with 40M parameters—an improvement over published result.

In summary, we (a) show that multi-aperture representation improves delineation of surface boundaries, and (b) propose an architecture that processes the multi-aperture representation by integrating Transformers and convolutional modules using 3D fusion blocks. The performance is also enhanced by incorporating the distance transform while reducing the model complexity. The organization of this paper is as follows. Section 2 provides a brief review of prior research in context. Section 3 outlines the methodology and its implementation. Section 4 outlines the results, visualization, and ablation studies and concludes the paper.

## 2. Related works

The early versions of ViTs [3, 22] were designed for the image classification tasks. ViT incorporates a self-attention mechanism, by means of pairwise similarities of input tokens, to capture long-range dependencies, which improved performance compared to ResNet [5] or vanilla UNet [13]. Recent studies [23, 24] have extended Transformer-based architectures to enhance the performance of medical segmentation tasks. For example, 3D TransUNet [11] integrates the Transformer blocks in the bottleneck to enhance performance. Other studies, such as nnFormer [4], Swin UNETR [2, 12], UNETR++ [6], and MIST [25] have also investigated alternative integrations of convolution and Transformers for improved performance. These models have effectively utilized course to fine representations through convolution or down sampling and integrated self-attention in unique ways. More recently, TransFuse [26] introduced a BiFusion module that fuses features from a shallow convnet-based encoder and the outputs of convolution and Transformer blocks in a unique way. This fusion block utilizes a diversity of dimensionality reduction techniques and ranks the feature maps' significance. Our goal is to build on prior arts, investigate the utility of multi-aperture representation, and to further improve delineation of surfaces in medical images.

## 3. Method

This section describes the proposed architecture, its implementation, and the customized loss function. Figure 1 illustrates the proposed architecture, for multi-aperture representation, that fuses the output of four Swin Transformers (e.g., Swin T1, Swin T2, Swin T3, Swin T4) with the Convolutional modules using the fusion blocks. While each Swin transformer computes global contextual relationships, the 3D convolution blocks compute local features from that representation but from a higher dimensional mapping.

### 3.1. Swin Encoders

The input data of the model is a 3D patch $x \in R^{H \times W \times D}$ that is randomly extracted from the raw data. This 3D patch is then processed by four Swin Transformer blocks [2], each with a different aperture, as shown in Figure 1. Each extracted patch $[H, W, D]$ serves as the input for the first Transformer block. Then, the center of the input for each block is selected iteratively as the input for the next block, i.e., the center of the first block at $[\frac{H}{2}, \frac{W}{2}, \frac{D}{2}]$ is selected as the input for the second block, the center of the input of the second block at $[\frac{H}{4}, \frac{W}{4}, \frac{D}{4}]$ is selected as input for the next block, etc.

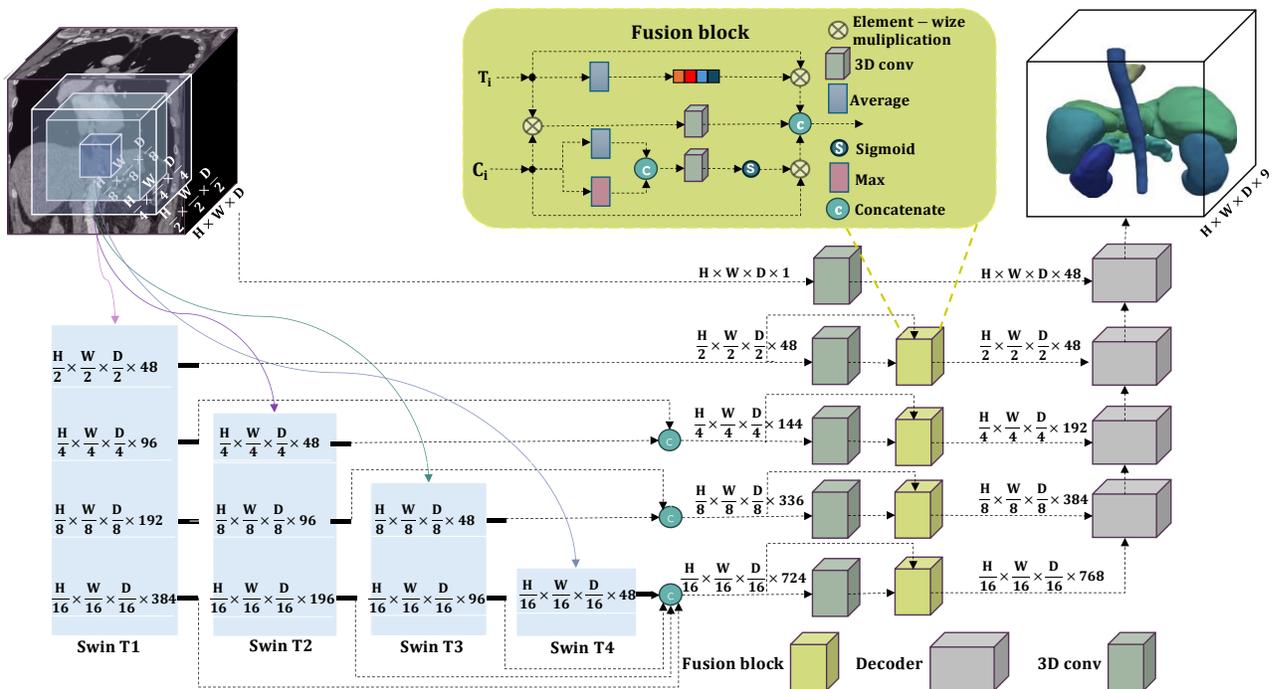

**Figure 1**: Proposed framework couples four Swin Transformer and their corresponding convolutional modules through fusion blocks in 3D. The output of four Swin Transformers, at different aperture, are fused with the output of convolutional modules for improved performance. The fusion block integrates the output of the transformer- ($T_i$) and its convolutional- ($C_i$) block using a diversity of summarization modules that include an SE, max pooling, and averaging blocks.

### 3.2. Fusion block

Fusion blocks, shown in Figure 1, integrate the global and local spatial features. Each fusion block concatenates the outputs of the (i) Transformer following the squeeze and excitation (SE) block [27] (color-coded block), (ii) element-wise multiplication of the output of the Transformer and Convolution, and (iii) convolutional module following Convolutional Block Attention Module (CBAM) [28]. Each of these outputs accentuates a specific aspect of the outputs of the Swin Transformers. For example, the SE block prioritizes the significance of channel-wise information, simple element-wise multiplication (e.g., Hadamard dot product) modulates the importance of features, and CBAM weights spatial similarities.

### 3.3. Loss function

The proposed loss function is computed by the combined errors computed from the DiceCE and the distance transform map. For each class$_i$, let $D_i$ be the absolute value of the signed distance transform of the predictor and let $S_i$ be the indicator function, set to one at the surface points and zero elsewhere. The distance transform is normalized between 0 and 1. With $\lambda$ as a hyperparameter, the combined loss function is defined as:

$$Loss(G, P) = DiceCE + \lambda \cdot \frac{\sum D_i \cdot S_i}{\sum D_i} \quad (1)$$

### 3.4. Implementation details

Experiments were conducted on a Linux Cluster Server equipped with 8 NVIDIA RTX 3080 GPUs, each with 12 GB of memory. All images were resized with the same physical dimensions of $1 \times 1 \times 1$ mmm, where each image stack varies from 85 to 198 slices. Subsequently, $128 \times 128 \times 128$ random patches were extracted to build a dataset with the augmentation techniques outlined in [2]. The model was trained for 300 epochs for patches that were randomly selected from 18 training images. Testing was performed on the remaining 12 images, and the best performing model, within the 300 epochs, was saved. The final performance is reported based on 5-fold cross validation. Learning is based on Adam optimization with a learning rate of $10^{-4}$, and a weight decay of $5 \times 10^{-5}$.

## 4. Results

The proposed model was compared to state-of-the-art (SOTA) CNN and Transformer models for medical image segmentation on the Synapse multi-organs dataset. This dataset comprises 30 abdominal CT scan cases with eight organs.

### 4.1. Quantitative evaluation

Table 1 shows the comparative performance of the proposed model against prior research, where our model produces a mean Dice score of 89.73±0.04 and HD95 7.31±0.02 with five-fold cross-validation. On average, these scores are better than the state-of-the-art studies 3D TransUNet and UNETR++. Moreover, out of eight classes of organs, we observed notable improvement in four: Spleen(spl), Right Kidney (Kid(R)), Left Kidney (Kid(L)), and Aorta (Aor). Finally, the proposed model has only 40M parameters, which, on average, performs better than prior research.

### 4.2. Visualization and comparison

Figure 2 provides a comparative visualization of the 3D segmentation results of four image samples with Swin UNETR, nnFormer, and 3D TransUNet. This result illustrates reduced spurious regions and continuity within each organ, which are visible for Swin UNETR and nnFormer. The differences between ours and 3D TransUNet can only be assessed at approximately 50% reduced complexity at this point.

| Methods | Avg Dice↑ | Avg HD95↓ | Par | Spl | Kid(R) | Kid(L) | Gal | Liv | Sto | Aor | Pan |
|---|---|---|---|---|---|---|---|---|---|---|---|
| MIST [1] | 86.92 | 11.07 | - | 92.83 | 93.28 | 92.54 | 74.58 | 94.94 | **87.23** | 89.15 | 72.43 |
| Swin UNETR [2] | 83.47 | 10.55 | 62M | 95.37 | 86.26 | 86.99 | 66.54 | 95.72 | 77.01 | 91.12 | 68.80 |
| nnFormer [4] | 86.57 | 10.63 | 150M | 90.51 | 86.25 | 86.57 | 70.17 | 96.84 | 86.83 | 92.04 | **83.35** |
| UNETR++ [6] | 87.22 | 07.53 | 42M | 95.77 | 87.18 | 87.54 | 71.25 | 96.42 | 86.01 | 92.52 | 81.10 |
| MISSFormer [7] | 81.96 | 18.20 | - | 91.92 | 82.00 | 85.21 | 68.65 | 94.41 | 80.81 | 86.99 | 65.67 |
| LeVit-UNet [10] | 78.53 | 16.84 | - | 88.86 | 80.25 | 84.61 | 62.23 | 93.11 | 72.76 | 87.33 | 59.07 |
| 3DTransUNet [11] | 88.10 | - | 81M | 93.39 | 87.47 | 85.76 | **81.15** | **97.34** | 85.31 | 92.97 | 81.76 |
| MFTC-Net | 89.17 | 07.47 | 40M | 95.98 | 92.66 | 93.30 | 76.41 | 96.61 | 85.05 | 92.62 | 80.77 |
| MFTC-Net +DistLoss | **89.73** | **07.31** | **40M** | **96.51** | **93.35** | **93.64** | 76.87 | 97.16 | 85.69 | **93.05** | 81.20 |

**Table 1:** Comparative evaluation indicates that the proposed model has a performance profile against published research at reduced complexity (Par for the number of parameters). Evaluation is performed against multiple organs (Gal for gallbladder, Kid(L) for left kidney, Kid(R) for right kidney, Pan for pancreas, Spl for spleen, Sto for stomach, Liv for Liver, and Aor for Aorta) and reported as the Mean Dice scores. The highest scores are emphasized in bold. The results are reported following five-fold cross-validation. The HD95 for 3DTransUNet has not been reported.

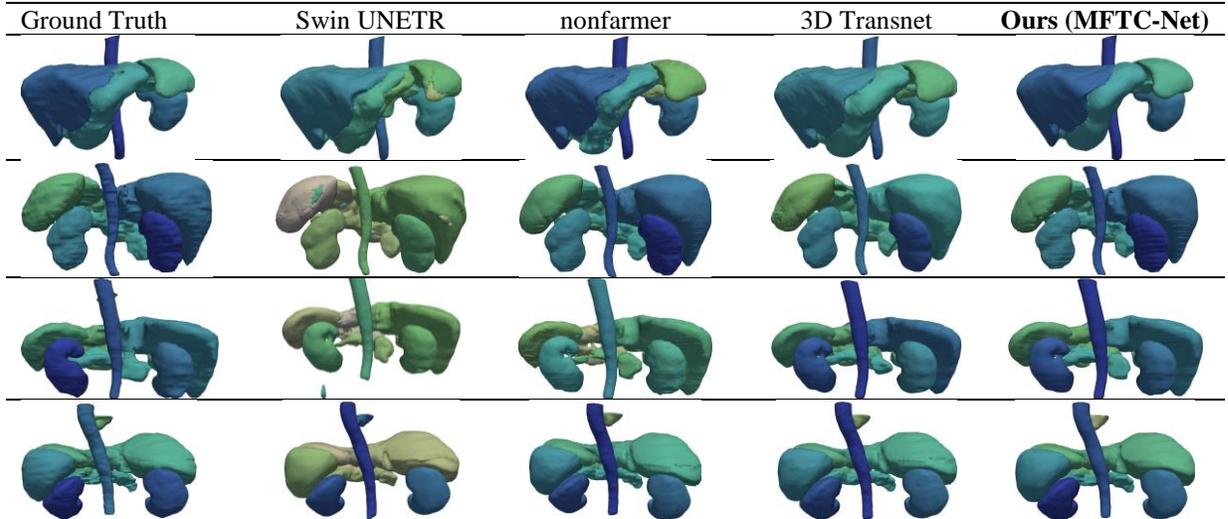

**Figure 2:** 3D Visualization of Segmentation Results on four samples of the Synapse multi-organ dataset indicate that our model has less spurious noise compared to swim UNETR and nonfarmer. The difference between ours and 3D TransUNet is minute and can only be assessed by the reduced number of parameters, i.e., 40 vs 81M. From left to right, the images depict the output of ground truth, Swin UNETR, nnFormer, 3D TransUNet, and our model (MFTC-Net), respectively.

### 4.3. Ablation study

Ablation studies include the effect of different loss functions and the number of Transformers and/or fusion blocks.

**4.2.1 Effect of alternative loss functions:** We examined the proposed framework using various objective function combinations, as detailed in Table 2. The dice loss, by itself, results in an average score of 89.04. In contrast, the use of a focal Dice loss [29] slightly improved the outcome, achieving an average dice score of 89.50. Furthermore, a hybrid approach that combined dice loss with binary cross entropy yielded a marginally better average dice score of 89.17. Most notably, introducing a novel term, DistLoss, into the mix with Dice loss and cross-entropy loss significantly enhanced performance, culminating in the highest average dice score of 89.73.

| Loss Function | Dice |
|---|---|
| $L_{dice}$ | 89.04 |
| $L_{focal}$ | 89.50 |
| $L_{dice} + L_{ce}$ | 89.17 |
| $L_{dice} + L_{ce} + DistLoss$ | **89.73** |

**Table 2:** Ablation study indicates an improved performance with the inclusion of DistLoss function.

**4.2.2 Effect of Transformers and Fusion Blocks:** Table 3 summarizes the impact of Transformer and fusion blocks on model performance and the total number of parameters.

| Blocks | Parameters | Dice |
|---|---|---|
| $T_1$ | 26M | 86.12 |
| $T_1 + T_2$ | 30M | 86.76 |
| $T_1 + T_2 + T_3$ | 34M | 87.39 |
| $T_1 + T_2 + T_3 + T_4$ | 36M | 88.75 |
| $T_1 + T_2 + T_3 + T_4 + Fusion$ | 40M | **89.73** |

**Table 3:** The ablation study indicates superior performance of multi-aperture Swin Transformer.

It is evident that the multi-aperture Transformers improve performance at a modest level by increasing the number of parameters.

### 5. Conclusions

We proposed a new framework, labeled as MFTC-Net, that includes four coupled Swin Transformers, at multiple apertures, with their corresponding convolutional blocks for medical image segmentation. The model was evaluated on the Synapse multi-organs dataset. In our framework, initially, Swin T1 receives an entire patch, while others receive a smaller aperture at the original resolution. As a result, the details are better preserved by maintaining the original image resolution. We also introduced a 3D fusion block for integrating the outputs of Transformers and convolution modules. Finally, we showed that a loss function based on the distance transform further enhances the overall performance based on detailed ablation studies. Our model outperformed published research on Synapse multi-organs datasets, achieving a new state of the art in terms of the mean Dice and mean HD95 scores while reducing the number of parameters.

**Acknowledgment:** The research was supported by a grant from NIH RO1CA279408.